%% file: paper1.tex
\documentclass[12pt]{article}

\usepackage{amsmath,amssymb,cite,float,fullpage,graphicx,slashed,authblk}
\usepackage{multirow}
\usepackage[T1]{fontenc}
\usepackage[table]{xcolor}
\usepackage{cancel}
\usepackage[]{hyperref}
\usepackage{url}
\hypersetup{
  bookmarks=true,         
  unicode=false,          
  pdftoolbar=true,        
 pdfmenubar=true,        
 pdffitwindow=true,     
 pdfstartview={FitH},    
 pdfsubject={GUT and Topological defects},   
 pdfnewwindow=true,      
 pdfcreator={RevTeX},
 colorlinks=true,       
 linkcolor=blue,          
 citecolor=blue,        
 filecolor=black,      
 urlcolor=blue,           
  }

\usepackage{hyperref}
\hypersetup{
linktocpage=true,
colorlinks=true,
linkcolor=blue,          
citecolor=mygreen,        
filecolor=blue,      
urlcolor=blue           
}  

\usepackage{fancyhdr}
\fancypagestyle{firstpage}{
  \fancyhf{} 
  \fancyhead[R]{\small\begin{tabular}{r}
  HU-EP-26/14-RTG
  \end{tabular}}
}

\definecolor{mygreen}{rgb}{0.0,0.75,0.0}

\usepackage{url}   

\usepackage{comment}

\usepackage{tikz}
\usetikzlibrary{arrows,positioning,matrix}
\usepackage{tikz-feynman}

\DeclareMathOperator{\tr} {tr}
\newcommand{\LL}{\mathrm{L}}
\newcommand{\RR}{\mathrm{R}}
\newcommand{\PW}{\mathrm{W}}
\newcommand{\PZ}{\mathrm{Z}}
\newcommand{\PH}{\mathrm{H}}
\newcommand{\Pt}{\mathrm{t}}
\newcommand{\MW}{M_\PW}
\newcommand{\MZ}{M_\PZ}
\newcommand{\MH}{M_\PH}
\newcommand{\mt}{m_\Pt}
\newcommand{\sw}{s_{\scriptscriptstyle\PW}}
\newcommand{\cw}{c_{\scriptscriptstyle\PW}}
\newcommand{\as}{\alpha_\mathrm{s}}
\newcommand{\at}{\alpha_\Pt}
\newcommand{\sigTa}{\Sigma_{\rm T(\alpha)}}
\newcommand{\sigTaas}{\Sigma_{\rm T(\alpha\as)}}
\newcommand{\sigTaaas}{\Sigma_{\rm T(\alpha^2\as)}}

\usepackage[normalem]{ulem}

\title{Third-order mixed electroweak-QCD corrections \\ to the W-boson mass prediction from the muon lifetime}
\author[1]{Ievgen Dubovyk}
\author[2]{Ayres Freitas}
\author[1]{Janusz Gluza}
\author[3]{Johann Usovitsch}
\affil[1]{\small Institute  of Physics, University of Silesia, Katowice, Poland}
\affil[2]{\small Pittsburgh Particle-physics Astro-physics \& Cosmology Center
(PITT-PACC), Department of Physics \& Astronomy, University of Pittsburgh, Pittsburgh, USA}
\affil[3]{Institut f\"ur Physik und IRIS Adlershof, Humboldt–Universität zu Berlin, 10099 Berlin,
Germany}
\date{} 

\begin{document}

\maketitle
\thispagestyle{firstpage}

\begin{abstract}
We present the calculation of the so far missing ${\cal O}(\alpha^2\alpha_\mathrm{s})$ corrections to the quantity $\Delta r$, which relates the Fermi constant to the W-boson mass, and enables precision predictions of the latter. While the ${\cal O}(\alpha^2\alpha_\mathrm{s})$ corrections from diagrams with two closed fermion loops are already known, we here focus on the subset with one closed fermion loop, which is a substantially more complex problem. The calculation has been carried out through a combination of analytical and numerical techniques for the three-loop integrals and the on-shell renormalization. The impact of the new corrections is numerically significant, raising the Standard Model prediction for the W-boson mass by more than 3~MeV.
\end{abstract}


\section{Introduction}

The charged gauge boson $W^\pm$ is a central element of the Standard Model (SM) of electroweak interactions
\cite{Salam:1968rm,Glashow:1961tr,Glashow:1970gm,Weinberg:1967tq},
mediating charged--current weak processes.
In particular, the purely leptonic muon decay
\begin{equation}
\mu^- \to \nu_\mu\, e^-\, \bar{\nu}_e,
\label{eq:muondecay}
\end{equation}
which proceeds via virtual $W^\pm$ exchange, has for decades provided a cornerstone of precision tests
of the SM, enabling a high-accuracy determination of the Fermi constant and,
indirectly, of the W-boson mass.

The current experimental precision for the W-boson mass measurement achieved by the CMS Collaboration reaches
$\delta \MW = 9.9~\text{MeV}$ ($\sim 0.012\%$) \cite{CMS:2024lrd}.
With an expected integrated luminosity at the level of
 $\mathcal{L}_{\text{HL-LHC}} \simeq 3~\text{ab}^{-1}$ \cite{ZurbanoFernandez:2020cco},
studies for precision measurements at the HL-LHC indicate that the total experimental uncertainty on $\MW$ could be reduced to
 $\delta \MW^{\text{HL-LHC}} \approx 5~\text{MeV}$ \cite{ATLAS:2018qzr}.
Future high-energy lepton colliders, most notably FCC-ee \cite{FCC:2018evy},
are expected to improve this precision substantially, targeting uncertainties
of $0.18~\text{MeV}$ (statistical) and $0.16~\text{MeV}$ (systematic) \cite{FCC:2025lpp}. 
Achieving this level of experimental accuracy will require commensurate advances
in theoretical predictions, in particular a precise description of the
$e^+e^- \to W^+W^-$ lineshape.

On the other hand, the presently available higher-order SM corrections to the muon decay, including NNLO and partial higher-order corrections, constrain the intrinsic
theoretical uncertainty of the W-boson mass to the level of approximately 4~MeV \cite{Awramik:2003rn}. 

In order to further improve the theoretical precision of the W-boson mass
and to meet the demands of future analyses of electroweak precision pseudo-observables (EWPOs),
we compute in this work the new
${\cal O}(N_f\,\alpha^2\as)$ contributions to muon decay within the full SM, where $N_f$ denotes the number of closed fermion loops in the contributing diagrams.
These corrections are of mixed electroweak--QCD origin and include genuine
three-loop amplitudes.
They retain the complete dependence on heavy SM particles,
including the top-quark mass $\mt$, the Higgs-boson mass $\MH$, and the gauge-boson
masses $\MZ$ and $\MW$.
Representative Feynman diagrams contributing to this class of corrections
are shown in Fig.~\ref{fig:diags}. \input{figures/figs_tikz.tex}

To set the stage, the muon lifetime $\tau_\mu$ can be conveniently expressed
within the Fermi theory in terms of the Fermi constant $G_\mu$ and QED radiative
corrections $\Delta q$, which are known up to third order \cite{Kinoshita:1958ru,Nir:1989rm,vanRitbergen:1999fi,Steinhauser:1999bx,Pak:2008qt,Fael:2020tow,Czakon:2021ybq},
\begin{equation}\label{eq:muon_lifetime}
    \tau_{\mu}^{-1}
    = \frac{G_\mu^2\, m_\mu^5}{192\pi^3}\,
      F\!\left(\frac{m_e^2}{m_\mu^2}\right)
      \left(1 + \Delta q\right),
\end{equation}
where
\begin{equation}
    F(x) = 1 - 8x - 12x^2 \ln x + 8x^3 - x^4
\end{equation}
is the phase-space factor resulting from the kinematic integration.

The experimentally measured muon lifetime is \cite{MuLan:2010shf,MuLan:2012sih}
\begin{equation}\label{eq:muon_lifetime_exp}
    \tau_{\mu}^{\rm exp}
    = 2.1969811 \pm 0.0000022~\mu{\rm s},
\end{equation}
which results in \cite{ParticleDataGroup:2024cfk}
\begin{equation}
    G_\mu = 1.1663785(6) \times 10^{-5}~\text{GeV}^{-2}. \label{eq:GFexp}
\end{equation}

Let us note in passing that the high-precision determination of the muon lifetime and Fermi coupling constant also provides a
powerful probe of physics beyond the SM. Examples include models with right-handed charged vector currents \cite{Czakon:2002wm,Czakon:1999ha}, heavy neutrinos and $Z'$ bosons  \cite{Atre:2009rg,Akhmedov:2013hec,Czakon:1999ha}, charged scalars \cite{Scheck:1977yg,Senjanovic:1978ee,Chankowski:2006hs}, and scalar and tensor Fermi interactions \cite{TWIST:2010uqs, ParticleDataGroup:2024cfk}. New physics effects can also be parametrized in terms of higher-dimensional, non-renormalizable operators in effective field theory descriptions \cite{Falkowski:2019hvp,deBlas:2017xtg}.
 
The precise value \eqref{eq:GFexp} for $G_\mu$ provides a basis for predicting the W-boson mass within the SM, according to the relation
\begin{equation}
    M_W^2 \left(1 - \frac{M_W^2}{M_Z^2}\right) = \frac{\pi \alpha}{\sqrt{2} G_\mu} \left(1 + \Delta r \right). \label{eq:deltar}
\end{equation}
Here $\Delta r = \Delta r(\alpha,\MW,\MZ,\MH,m_f)$ collects all possible SM radiative corrections \cite{Sirlin:1980nh}. Over a period of many years, NLO
\cite{Sirlin:1980nh} and NNLO electroweak corrections\cite{Freitas:2000gg,Awramik:2002wn,Awramik:2002vu,Onishchenko:2002ve,Awramik:2003ee}, as well as mixed electroweak-QCD NNLO corrections \cite{Djouadi:1987di,Kniehl:1989yc,Kniehl:1991gu,Djouadi:1993ss} have been computed. In addition, higher-order corrections enhanced by the top Yukawa coupling $y_\Pt$ are known at ${\cal O}(\at\as^2)$ \cite{Avdeev:1994db,Chetyrkin:1995ix,Chetyrkin:1995js}, ${\cal O}(\at^2\as)$ and ${\cal O}(\at^3)$ \cite{vanderBij:2000cg,Faisst:2003px}, and ${\cal O}(\at\as^3)$ \cite{Schroder:2005db,Chetyrkin:2006bj,Boughezal:2006xk}, where $\at \equiv y_\Pt^2/(4\pi)$. More recently, leading fermionic third-order corrections of order ${\cal O}(N_f^3\alpha^3)$ and  ${\cal O}(N_f^2\alpha^2\as)$ \cite{Chen:2020xzx,Chen:2020xot} have become available. The corrections listed in this paragraph have been implemented in the {\tt GRIFFIN} package \cite{Chen:2022dow}.
Recent developments in the three-loop QED and $\overline{\text{MS}}$ scheme  corrections to the Fermi decay constant are 
reported in \cite{Duhr:2024bzt} and \cite{Martin:2025cas}, respectively.

The ${\cal O}(N_f\,\alpha^2\as)$ results presented in this paper are distinct from the ${\cal O}(N_f^2\,\alpha^2\as)$ effects discussed in Ref.~\cite{Chen:2020xot}, and they require the calculation of genuine three-loop amplitudes. We employ the on-shell (OS) renormalization scheme, which allows us to present the result directly in terms of experimentally determined input parameters, such as particle masses\footnote{A calculation of $\Delta r$ in the $\overline{\text{MS}}$ scheme was recently accomplished in Ref.~\cite{Martin:2025cas}.}. For the OS counterterms, it is necessary to evaluate three-loop self-energies at non-zero momentum.


\section{Description of the calculation}
\label{sec:calc}

Our objective is the computation of genuine  ${\cal O}(N_f\,\alpha^2\as)$  three-loop corrections in the full SM, involving several massive particles propagating inside the loops. Such calculations are highly demanding and require careful and consistent bookkeeping to enable non-trivial cross-checks at intermediate stages of the computation. Therefore, for most building blocks of the total result, we have performed two entirely independent calculations involving the entire tool chain from diagram generation, algebraic manipulations, integration-by-parts (IBP) reduction, and evaluation of master integrals (MIs).

Owing to the mass hierarchy $m_\mu \ll \MW$, the relevant matrix element for $\mu^- \to \nu_\mu\, e^-\, \bar{\nu}_e$ can be evaluated in the limit where all external momenta and all lepton masses are set to zero. A priori, this simplification is not justified for infrared (IR) divergent diagrams with photon exchange between the external charged leptons. However, these IR-sensitive contributions cancel when matching to the muon-decay matrix element in the Fermi theory, as described in section~\ref{sec:fermi} below. 

Furthermore, we also neglect light fermion masses, $m_{f\neq t}$, inside the loops and set the CKM matrix equal to the unit matrix, both of which are numerically good approximations. By default, we use dimensional regularization for the regularization of ultra-violet (UV) divergences and a photon mass for the regularization of IR divergences.

Feynman amplitudes are generated with \texttt{FeynArts~3.11} \cite{Hahn:2000kx}, which is a package in \texttt{Mathematica} \cite{Mathematica}, and with \texttt{DiaGen} \cite{Czakon:DiaGen} and \texttt{FORM} \cite{Kuipers:2012rf,Ruijl:2017dtg,Vermaseren:2000nd}. For \texttt{DiaGen}, we prepared custom electroweak model files following Ref.~\cite{Denner:2019vbn}. The Lorentz and Dirac algebra is performed using \texttt{FeynCalc~9} \cite{Shtabovenko:2016sxi,Mertig:1990an} and \texttt{FormCalc} \cite{Hahn:2016ebn,Hahn:1998yk}, and cross-checked with independent in-house implementations in \texttt{Mathematica} and \texttt{FORM}. The IBP reduction of Feynman integrals to MIs is carried out with \texttt{Kira~3} \cite{Lange:2025fba,Klappert:2020nbg,Maierhofer:2018gpa,Maierhofer:2017gsa,Klappert:2019emp,Klappert:2020aqs} and \texttt{FIRE~5} \cite{Smirnov:2014hma,Smirnov:2013dia,Smirnov:2008iw} and in-house programs written in \texttt{FORM}. The resulting MIs are then computed numerically with the help of \texttt{AMFlow} \cite{Liu:2022chg} and \texttt{TVID2} \cite{Bauberger:2019heh,Bauberger:2017nct}. Some MIs have been cross-checked against analytical results in the literature \cite{Schroder:2005va,Bytev:2009kb}, performing expansions in $\epsilon = (4-d)/2$ with \texttt{HypExp~2} \cite{Huber:2007dx}, where $d$ is the space-time dimension in dimensional regularization.

The on-shell renormalization also requires the evaluation of ${\cal O}(N_f\,\alpha^2\as)$ self-energy at non-zero momentum. Owing to the complexity of these integrals, we have carried out only one full calculation, using \texttt{Kira} and \texttt{AMFlow}
\footnote{
Recent computations of the full two-loop corrections to $Z$-boson production and decay \cite{Dubovyk:2016aqv,Dubovyk:2018rlg,Dubovyk:2019szj} relied on numerical evaluations based on sector decomposition (SD) \cite{Binoth:2000ps,Borowka:2017idc,Borowka:2018goh,Smirnov:2008py,Smirnov:2015mct} and Mellin–Barnes representations \cite{Gluza:2007rt,Czakon:2005rk,Usovitsch-phd:2018qmt,Dubovyk:2016ocz,Usovitsch:2018shx,Dubovyk:2016aqv}. While powerful, these methods are computationally expensive and do not always converge to the accuracy required for phenomenological applications, rendering a straightforward extension to higher loop orders and/or additional external legs impractical. At the three-loop level, the number of Feynman diagrams typically increases by roughly two orders of magnitude \cite{Blondel:2018mad}. Based on previous experience \cite{Dubovyk:2018rlg}, one expects precision losses of up to five digits due to numerical cancellations between individual loop integrals, implying that at least 8–10 digits of precision are often required. At present, achieving this level of accuracy on a large scale is not feasible with existing SD- and MB-based numerical tools. Instead, new methods based on series solutions of differential equations have proven superior for cutting-edge loop calculations \cite{Liu:2017jxz,Moriello:2019yhu,Hidding:2020ytt,Dubovyk:2022frj,Prisco:2025wqs}.}. However, a subset of self-energy integrals has been cross-checked 
using \texttt{pySecDec} \cite{Borowka:2017idc,Heinrich:2021dbf}, \texttt{FIESTA} \cite{Smirnov:2008py,Smirnov:2015mct,Smirnov:2021rhf}, \texttt{AmpRed} \cite{Chen:2024xwt}, and \texttt{TVID2}.
For topology identification we used  \texttt{TopoID} \cite{Grigo:2014oqa,githubtopoid} (for the original concept, see \cite{Pak:2011xt}). 

In addition to genuine three-loop contributions, we also need to evaluate two-loop diagrams with one-loop counterterm insertions and one-loop diagrams with two-loop counterterm insertions. These have been handled with the help of the same packages listed above, together with in-house code for the work flow and bookkeeping.

For the final result, a cancellation of the coefficients of the UV-divergent $\epsilon^{-3}$, $\epsilon^{-2}$ and $\epsilon^{-1}$ terms at 19, 15 and 8 digits, respectively, has been achieved, which is an important validation of the technical calculation and the renormalization procedure.

\subsection{Renormalization \label{sec:renorm}}

The renormalization is performed in the on-shell (OS) scheme, which uses fermion masses, W/Z/H boson masses, and the electromagnetic coupling as independent parameters. Concretely, the on-shell masses are defined via the location of the pole of their radiatively corrected propagators. For particles with non-negligible decay widths, the propagator pole is complex, which affects the mass counterterms at NNLO and beyond \cite{Freitas:2002ja}. For the purposes of the present work, this needs to be considered for the W and Z boson masses, $M_{\PW,\PZ}$, and the top-quark mass, $m_\Pt$. More detailed expressions for the mass counterterms can be found in the appendix.

It should be noted that $M_{\PW,\PZ}$ defined in this way differ numerically from the mass values reported in experimental analyses. The relation between them is given by \cite{Bardin:1988xt}
\begin{align}
\begin{aligned}
M_\PZ &= M_\PZ^{\rm exp}[1+(\Gamma_\PZ^{\rm exp}/M_\PZ^{\rm exp})^2]^{-1/2} 
 \approx M_\PZ^{\rm exp} - 34\,\text{MeV}, \\
M_\PW &= M_\PW^{\rm exp}[1+(\Gamma_\PW^{\rm exp}/M_\PW^{\rm exp})^2]^{-1/2} 
 \approx M_\PW^{\rm exp} - 27\,\text{MeV}.
\end{aligned} \label{mztrans}
\end{align}
For the top-quark mass, the analogous shift is of the order of 5~MeV and negligible for practical purposes.

For the charge renormalization, one needs to carry out the derivative of the photon self-energy at zero momentum, $\frac{\partial}{\partial(p^2)}\Sigma^{\gamma\gamma}(p^2=0)$. For massive particles in the loops (W/Z/t), this is accomplished by the following procedure. Let us define a generic three-loop self-energy integral according to
\begin{align}
    &I(\nu_1,\nu_2,...;p^2) \equiv \int \frac{d^dq_1\,d^dq_2\,d^dq_3}{D_1^{\nu_1}D_2^{\nu_2}D_3^{\nu_3}D_4^{\nu_4}D_5^{\nu_5}D_6^{\nu_6}D_7^{\nu_7}D_8^{\nu_8}D_9^{\nu_9}}, \\
    &\begin{aligned}
    &D_1 = q_1^2-m_1^2, & &D_4 = q_2^2-m_4^2, & &D_7 = q_3^2-m_7^2, \\
    &D_2 = (q_1+p)^2-m_2^2, & &D_5 = (q_2+p)^2-m_5^2, & &D_8 = (q_3+p)^2-m_8^2, \\
    &D_3 = (q_2-q_1)^2-m_3^2, & &D_6 = (q_3-q_1)^2-m_6^2, & &D_9 = (q_3-q_2)^2-m_9^2.
    \end{aligned}
\end{align}
Then
\begin{align}
    &\frac{\partial}{\partial(p^2)}I(\dots;p^2=0) = \frac{1}{2d} \; \frac{\partial^2}
  {\partial p_\mu \partial p^\mu} I(\dots; p^2) \bigg|_{p^2=0} \nonumber \\
&\quad = \frac{2}{d} \Bigl[ (1+\nu_2+\nu_5+\nu_8-d/2)\,
  [ \nu_2 I(\nu_2+1) + \nu_5 I(\nu_5+1) + \nu_8 I(\nu_8+1) ] \nonumber \\
& \quad\qquad + m_2^2\nu_2(\nu_2+1) I(\nu_2+2) + m_5^2\nu_5(\nu_5+1) I(\nu_5+2) + m_8^2\nu_8(\nu_8+1) I(\nu_8+2) \\
& \quad\qquad + \nu_2\nu_5 \left( (m_2^2-m_3^2+m_5^2) I(\nu_2+1,\nu_5+1)
  - I(\nu_2+1,\nu_3-1,\nu_5+1) \right) \nonumber \\
& \quad\qquad + \nu_2\nu_8 \left( (m_2^2-m_6^2+m_8^2) I(\nu_2+1,\nu_8+1)
  - I(\nu_2+1,\nu_6-1,\nu_8+1) \right) \nonumber \\
& \quad\qquad + \nu_5\nu_8 \left( (m_5^2-m_9^2+m_8^2) I(\nu_5+1,\nu_8+1)
  - I(\nu_5+1,\nu_8+1,\nu_9-1) \right) \Bigr]_{p^2=0},\nonumber 
\end{align}
\emph{i.e.} one obtains a linear combination of 3-loop vacuum integrals with propagators raised to higher powers, which can be reduced to MIs by using IBP identities. 

For only massless bosons and light quarks in the loops, the limit $m_q \to 0$ of \mbox{$\frac{\partial}{\partial(p^2)}\Sigma^{\gamma\gamma}(p^2=0)$} is singular. This indicates the presence of non-perturbative hadronic corrections at low $p^2$, which are typically absorbed into a running electromagnetic coupling, $\alpha(p^2)$, according to: 
\begin{align}
    &\alpha(\MZ^2) \equiv \frac{\alpha(0)}{1-\Delta\alpha}, &
    &\Delta\alpha = \frac{\partial}{\partial(p^2)}\Sigma^{\gamma\gamma}_{\rm lq,noweak}(p^2=0) - \frac{\Sigma^{\gamma\gamma}_{\rm lq,noweak}(\MZ^2)}{\MZ^2}.
\end{align}
Here ``lq,noweak'' refers to contributions without top quark and massive W/Z bosons in the loops. $\Sigma^{\gamma\gamma}_{\rm lq,noweak}(\MZ^2)$ can be obtained from the ${\cal O}(N)$ part of eq.~(2.7) of Ref.~\cite{Gorishnii:1991hw} with straightforward coupling replacements.

\subsection{Treatment of \boldmath $\gamma_5$}
\label{sec:gam5}

A subset of the vertex corrections with a fermion subloop receive non-vanishing contributions from traces including $\gamma_5$,
\begin{align}
\tr\{\gamma^\alpha\gamma^\beta\gamma^\gamma\gamma^\delta\gamma_5\} = -4i\,\epsilon^{\alpha\beta\gamma\delta}, \label{eq:tr5}
\end{align}
where $\epsilon^{\alpha\beta\gamma\delta}$ is the fully antisymmetric Levi-Civita tensor. Examples of such diagrams are shown in Fig.~\ref{fig:diags}~(d,e), and they involve ${\cal O}(\as)$ corrections to PCAC-violating fermion triangles (PCAC = partially conserved axial-vector current). According to a famous theorem by Adler and Bardeen \cite{Adler:1969er}, higher-order corrections to PCAC violation are absent, which would imply that the sum over diagrams of the form in Fig.~\ref{fig:diags}~(d,e), \emph{i.e.} diagrams with gluonic corrections inside a quark sub-loop triangle, is zero. However, the Adler--Bardeen theorem assumes that all fermion propagators in a given triangle diagram have the same mass, which is not the case in our calculation, where both top and bottom quarks appear in the same diagram.

As a result, we can receive a non-zero contribution from these diagrams, and an explicit calculation is necessary. Due to symmetry and renormalization group arguments, this contribution is required to be UV finite \cite{Costa:1977pd}, but the results of individual diagrams can be divergent and must be regularized. It is well known that standard dimensional regularization is incompatible with eq.~\eqref{eq:tr5}. Therefore, we instead resort to Pauli-Villars (PV) regularization \cite{Pauli:1949zm} for this subset of vertex diagrams, which corresponds to the following replacement for a gluon propagator,
\begin{align}
    \frac{1}{k^2} \to \frac{1}{k^2} -\frac{1}{k^2-\Lambda^2}\,, \label{eq:PV}
\end{align}
and taking the limit $\Lambda \to \infty$ in the end. While PV regularization breaks gauge-invariance for a non-Abelian gauge theory, this approach is justified here, since the relevant diagrams only contain Abelian-type gluon couplings. Note that the replacement \eqref{eq:PV} must also be applied inside the mass counterterm in Fig.~\ref{fig:diags}~(e). In practice, we have carried out the calculation for two large values of $\Lambda$, $\Lambda_1 = 5\times 10^7$~GeV and $\Lambda_2 = 5\times 10^8$~GeV, and verified that the results are consistent to more than 8 digits accuracy.

\subsection{Matching with Fermi theory \label{sec:fermi}}

After computing all contributing ${\cal O}(N_f\,\alpha^2\as)$ corrections to muon decay in the SM, one must perform the matching to the Fermi theory in order to remove IR singularities and extract a finite result for $\Delta r$. Examples of IR-singular contributions are the box diagrams in Fig.~\ref{fig:diags}~(a,b,c) with photons, as well as photonic contributions to the field renormalization counterterms of the external muon and electron. More specifically, the matching requires the subtraction of virtual QED corrections encompassed in $\Delta q$ in eq.~\eqref{eq:muon_lifetime}. Sample diagrams are shown in Fig.~\ref{fig:diags}~(h,i). These corrections are computed in the Fermi model + QED + QCD.

When evaluating radiative corrections to muon decay in the Fermi model, it is advantageous to apply a Fierz transformation \cite{Berman:1962uvx},
\begin{align}
    \frac{G_\mu}{\sqrt{2}}\,\bigl[\overline{\psi}_{\nu_\mu}\gamma_\mu P_\LL \psi_\mu\bigr]\bigl[\overline{\psi}_e \gamma^\mu P_\LL \psi_{\nu_e}\bigr] \;\to\; \frac{G_\mu}{\sqrt{2}}\,\bigl[\overline{\psi}_e\gamma_\mu P_\LL \psi_\mu\bigr]\bigl[\overline{\psi}_{\nu_\mu} \gamma^\mu P_\LL \psi_{\nu_e}\bigr],
\end{align}
where $P_{\LL} = (1- \gamma_5)/2$. In this way, the QED interactions are confined to one of the two Fermion lines, and the Dirac algebra is the same as for a vertex form factor. To avoid complications with evanescent operators in dimensional regularization, we calculate the corrections to $\Delta q$ in four dimensions using PV regularization\footnote{An alternative approach using dimensional regularization at two-loop order has been presented in Refs.~\cite{Awramik:2002vu,Awramik:2003ee}, but it is unclear whether it generalizes to higher orders.}. It turns out that the sum of these virtual corrections is exactly zero at the perturbative order considered here.


\section{Numerical results and conclusion \label{sec:results}}
\label{sec:results}

By combining all the ingredients discussed in the previous section, a numerical result for the ${\cal O}(N_f\,\alpha^2\as)$ correction to $\Delta r$ is obtained. We use the following input parameters:
\begin{align}
    &\MZ = 91.1876~\text{GeV}, &
    &\MW = 80.385~\text{GeV}, &
    &\MH = 125.1~\text{GeV}, &
    &\mt = 173.2~\text{GeV}.
    \label{eq:initval}
\end{align}
In principle, input values for the masses should be translated according to \eqref{mztrans}. However, this translation can be regarded as a higher-order correction on top of the NNNLO corrections discussed here, and thus it is negligible for practical purposes. We will present a more detailed numerical analysis, including the dependence on the input parameters, in a future publication.

Furthermore, we note that $\Delta\alpha$ is not needed as an input parameter for the ${\cal O}(N_f\,\alpha^2\as)$ corrections. The reason is that $\Delta\alpha$ is a non-perturbative parameter which absorbs all higher-order QCD corrections. Since our new result is the ${\cal O}(\as)$ correction to the previously known ${\cal O}(N_f\,\alpha^2)$ result, the dependence on $\Delta\alpha$ of the latter already captures the running of the electromagnetic coupling at ${\cal O}(N_f\,\alpha^2\as)$.

With these input values, the  correction to $\Delta r$ is found to be
\begin{align}
    \Delta r^{(N_f\alpha^2\as)} = -1294.9\, C_F \, \Bigl(\frac{\alpha}{\pi}\Bigr)^2 \frac{\as}{\pi}. \label{drrestot}
\end{align}
This result includes the top-Yukawa-enhanced contribution to the $\rho$ parameter previously computed in Ref.~\cite{Faisst:2003px}. To isolate the new contribution beyond $\Delta\rho$, we therefore evaluate the difference
\begin{align}
    \Delta r^{(N_f\alpha^2\as)}_{\rm beyond\,\Delta\rho} = \Delta r^{(N_f\alpha^2\as)} + \frac{\cw^2}{\sw^2}\Delta\rho^{(\alpha^2\as)},
\end{align}
using the results of Ref.~\cite{Faisst:2003px} for $\Delta\rho^{(\alpha^2\as)}$. For consistency, we parametrize the latter in terms of $X_t = \frac{\alpha \mt^2}{16\pi \sw^2\MW^2}$ rather than in terms of $x_t = \frac{G_\mu\mt^2}{8\sqrt{2}\pi^2}$. We obtain
\begin{align}
    \frac{\cw^2}{\sw^2}\Delta\rho^{(\alpha^2\as)} &= 549.9\, C_F \, \Bigl(\frac{\alpha}{\pi}\Bigr)^2 \frac{\as}{\pi}, \label{drhores} \\
    \Delta r^{(N_f\alpha^2\as)}_{\rm beyond\,\Delta\rho} &=-745.0\, C_F \, \Bigl(\frac{\alpha}{\pi}\Bigr)^2 \frac{\as}{\pi} = -2.0 \times 10^{-4},
\end{align}
where in the last step we have inserted $\alpha = 1/137.037$ and $\as = 0.118$.

To estimate the shift, $\Delta\MW$, on the predicted value of the W-boson mass, one can expand \eqref{eq:deltar} to linear order in $\Delta\MW$ and $\Delta r$, yielding
\begin{align}
    \Delta\MW = \frac{\pi\alpha\MZ^2}{2\sqrt{2}G_\mu\MW(\MZ^2-2\MW^2)}\Delta r^{(N_f\alpha^2\as)}_{\rm beyond\,\Delta\rho} = + 3.14 \text{ MeV}. \label{mwres}
\end{align}
This numerical impact on the W-boson mass is quite significant compared to other known partial three-loop contributions, see Tab.~\ref{tab:orders}. Previously, the theory uncertainty due to the then unknown ${\cal O}(\alpha^2\as)$ corrections was estimated to be 3~MeV \cite{Awramik:2003rn}, which is similar in size to the actual result in eq.~\eqref{mwres}.

\begin{table}[tb]
\centering
\begin{tabular}{lrr}
\hline\hline
 & $\Delta r$ [$10^{-4}$] & $\Delta M_W$ [MeV] \\
\hline
$\ \mathcal{O}(\alpha)$ & $+296.8$\phantom{0} &   $-441.5$\phantom{0} \\
$\ \mathcal{O}(\alpha\as,\,\alpha\as^2)$ & $+42.2$\phantom{0}  & $-73.0$\phantom{0}  \\
$\ \mathcal{O}(\alpha^2)$ & $+29.0$\phantom{0} & $-49.4$\phantom{0} \\
$\ \mathcal{O}(\at^2\as,\, \at^3,\, \at\as^3)$ 
& $-0.30$  & $+0.50$  \\
$\ \mathcal{O}(N_f^3\alpha^3,\, N_f^2\alpha^2\as)$ & $-0.89$  & $+1.46$  \\
$\ \mathcal{O}(N_f\alpha^2\as)$ [\rm{this work}]& $-2.01$  &  $+3.14$ \\
\hline\hline
\end{tabular}
\caption{Perturbative contributions for the muon decay parameter $\Delta r$ in the SM and the resulting $M_W$ change. The rows above the last one have been evaluated with {\tt GRIFFIN} \cite{Chen:2022dow}.}
\label{tab:orders}
\end{table}

It may be noted that the full ${\cal O}(\alpha^2\as)$ correction in eq.~\eqref{drrestot} is significantly larger than the contribution from $\Delta\rho$ in eq.~\eqref{drhores}, suggesting limitations of the reliability of the large-$\mt$ expansion used in Ref.~\cite{Faisst:2003px}. Since the MIs are evaluated numerically in our calculation, we cannot extract this leading ${\cal O}(\at^2\as)$ term. However, to investigate this matter further, we have computed the finite part of the combination
\begin{align}
    \Delta\bar\rho^{(\alpha^2\as)} \equiv \frac{\sigTaaas^\PZ(0)}{\MZ^2} - \frac{\sigTaaas^\PW(0)}{\MW^2}\,,
\end{align}
where $\Sigma_{\rm T}^{\PZ,\PW}(p^2)$ are the transverse Z/W-boson self-energies. $\Delta\bar\rho^{(\alpha^2\as)}$ contains all orders in $\mt^{-2}$, whereas $\Delta\rho^{(\alpha^2\as)}$ captures only the leading $\mt^4$ term.
For the input values in eq.~\eqref{eq:initval} we find $\Delta\bar\rho^{(\alpha^2\as)}/\Delta\rho^{(\alpha^2\as)} \approx 1.81$, where for $\Delta\rho^{(\alpha^2\as)}$ we used the results from Ref.~\cite{Faisst:2003px}. This again indicates that the terms beyond the leading $\mt^4$ term are numerically large. 

When increasing the top-quark mass by a factor 2, to $\mt=346.4$~GeV, the ratio changes to $\Delta\bar\rho^{(\alpha^2\as)}/\Delta\rho^{(\alpha^2\as)} \approx 1.29$, \emph{i.e} the relative difference is reduced by a factor of about 3. This is consistent with the expectation that the difference between $\Delta\bar\rho^{(\alpha^2\as)}$ and $\Delta\rho^{(\alpha^2\as)}$ is dominated by the formally sub-leading $\mt^2$ term, which parametrically is reduced by a factor 4 compared to the leading $\mt^4$ term when doubling the value of $\mt$. In practice, this simple factor 4 is modulated by higher-order terms and logarithmic terms in the $\mt^{-2}$ series.

We therefore conclude that the leading ${\cal O}(\at^2\as)$ term in a large-$\mt$ expansion is not a good approximation of the full ${\cal{O}}(\alpha^2\as)$ result.
This is reminiscent of a situation that was encountered for the ${\cal O}(\alpha^2)$ corrections \cite{Degrassi:1996mg}.

\begin{table}[tb]
\centering
\begin{tabular}{lc}
\hline\hline \\[-3ex]
Contribution & 
\parbox{4cm}{Correction to $\Delta r$, \newline in units of $C_F \, (\frac{\alpha}{\pi})^2 \frac{\as}{\pi}$} \\[1.5ex]
\hline
IR-finite boxes & 154.2 \\
QED boxes, with Fermi
model contribution subtracted & $-1.6 - 8.4\,\ln\frac{m_\gamma}{\rm GeV}$    \\
Traces with $\gamma_5$ in vertex diagrams 
& 2.8 \\
1-particle reducible contributions  &  $9.8 + 12.0\,\ln\frac{m_\gamma}{\rm GeV}$ \\
\hline
$\mathcal{O}(\alpha^2\as)$ weak mixing angle counterterm (finite part) & $-1229.1$  \\
All remaining contributions (finite part) & $-231.0 - 3.7\,\ln\frac{m_\gamma}{\rm GeV}$ \\
\hline
Total & $-1294.9$  \\
\hline\hline
\end{tabular}
\caption{Breakdown of individual contributions to the $\mathcal{O}(N_f\alpha^2\as)$ correction of $\Delta r$.  }
\label{tab:parts}
\end{table}

Table~\ref{tab:parts} shows the numerical impact of different individual contributions to the $\mathcal{O}(N_f\alpha^2\as)$ contribution. The first four rows are subsets that are individually UV-finite. Some sample diagrams for the IR-finite boxes are shown in Fig.~\ref{fig:diags}~(a,b,c) with at least one Z-boson. The next row in the table refers to boxes with photons, see again Fig.~\ref{fig:diags}~(a,b,c) for examples, matched to the Fermi theory as described in section~\ref{sec:fermi}. This contribution is IR-divergent, due to the lepton field renormalization counterterms in the Fermi model, and thus depends on the photon mass regulator $m_\gamma$. Its numerical impact is rather small, since there are no particular enhancements from the top Yukawa coupling or combinatorial factors. The third row in the table (``Traces with $\gamma_5$ ...'') accounts for the PCAC-violating terms discussed in section~\ref{sec:gam5}, which are both finite and gauge-invariant. The fourth row shows the result from one-particle reducible diagrams, which contain separate ${\cal O}(\alpha)$ and ${\cal O}(\alpha\as)$ sub-diagrams connected by a W propagator. This contribution has an IR-divergence stemming from lepton and W-boson field renormalization counterterms.

From the remaining contributions to $\Delta r^{(N_f\alpha^2\as)}$, the 3-loop weak mixing counterterm, $\delta\sw{}_{(\alpha^2\as)}$, is separately shown in the fifth row of Tab.~\ref{tab:parts}. This counterterm is not UV-finite, and only the finite $\epsilon^0$ part is given in the table. It is the numerically dominant contribution to the overall result for $\Delta r^{(N_f\alpha^2\as)}$, which can be traced to the $\Delta\bar\rho$-like correction contained in $\delta\sw{}_{(\alpha^2\as)}$.  

To summarize, in this work we calculated the 3-loop 
${\cal O}(N_f\,\alpha^2\as)$ contribution to muon decay within the full SM, matched to the Fermi model prediction. The result produces a positive shift of 3.14 MeV in the W-boson mass prediction within the SM and represents an important step towards precision studies of the electroweak observables at HL-LHC and at future $e^+e^-$ colliders. 
A reliable assessment of how the present result reduces the theoretical uncertainty in $\Delta M_W$ relative to the previous estimate of 4 MeV requires a more detailed study and is left for future work.
With the present result, the only missing third-order correction to the W-boson mass prediction from the muon lifetime is the purely electroweak contribution $\mathcal{O}(\alpha^3)$. 

\section*{Acknowledgments}
The authors like to recognize contributions by  Krzysztof Grzanka for the development of some numerical packages and initial tests.
J.U.\ would like to thank Micha\l\ Czakon for sharing his program \texttt{Diagen}. 
This work has been supported in part by the Polish National Science Centre (NCN) under Maestro grant No.~2023/50/A/ST2/00224, the U.S.~National Science Foundation under grant nos.~PHY-2112829 and PHY-2412696, and the Research Excellence Initiative of the University of Silesia in Katowice, by the Deutsche Forschungsgemeinschaft
(DFG, German Research Foundation) Projektnummer 417533893/GRK2575 ``Rethinking Quantum Field Theory'' and by the European Union through the 
European Research Council under grant ERC Advanced Grant 101097219 (GraWFTy).


\appendix
\section*{Appendix: Expressions for higher-order counterterms}

The renormalization scheme used in this calculation introduces bosonic and fermion mass counterterms,
\begin{align}
M_B^2 &\to M_B^2 + \delta M_B^2, & m_f &\to m_f + \delta m_f,
\end{align}
with $B=\PW,\PZ,\PH$ and $f=\Pt$ (all masses for lighter fermions are neglected). In addition, there are counterterms for charge and field renormalization:
\begin{align}
    e &\to Z_e e, & Z &\to \sqrt{Z^{\PZ\PZ}}\, Z + \tfrac{1}{2}\delta Z^{\PZ\gamma}\, A, & f_\LL &\to \sqrt{Z^{f\LL}}\, f_\LL, \\
    W &\to \sqrt{Z^\PW}\,W, & A &\to \tfrac{1}{2}\delta Z^{\gamma\PZ}\,Z + \sqrt{Z^{\gamma\gamma}}\, A, & f_\RR &\to \sqrt{Z^{f\RR}}\, f_\RR.
\end{align}
Field renormalization constants for other fields are not explicitly needed in this calculation.
Lowest-order expressions for the counterterms in the OS scheme can be found, \emph{e.g.}, in Ref.~\cite{Denner:2019vbn}. Here we report results that require non-trivial extensions to account for higher-order effects.

Since the widths of the W boson, Z boson and top quark are non-negligible, their masses are defined via the complex pole of the propagator. For the W- and Z-boson mass counterterms, the relevant expressions can be derived straightforwardly from Ref.~\cite{Freitas:2002ja}:
\begin{align}
    \delta M^2_{\PW(\alpha^2\as)} &= \Re\{\sigTaaas^\PW(\MW^2)\} - \delta M^2_{\PW(\alpha)} \, \delta Z^\PW_{(\alpha\as)} - \delta M^2_{\PW(\alpha\as)} \, \delta Z^\PW_{(\alpha)} \notag \\
    &\quad + \Im\{\sigTa^{\PW\prime}(\MW^2)\}\, \Im\{\sigTaas^\PW(\MW^2)\} + \Im\{\sigTaas^{\PW\prime}(\MW^2)\}\, \Im\{\sigTa^\PW(\MW^2)\}\,,  \label{eq:dmw} \\[1ex]
    \delta M^2_{\PZ(\alpha^2\as)} &= \Re\{\sigTaaas^\PZ(\MZ^2)\} - \delta M^2_{\PZ(\alpha)} \, \delta Z^{\PZ\PZ}_{(\alpha\as)} - \delta M^2_{\PZ(\alpha\as)} \, \delta Z^{\PZ\PZ}_{(\alpha)} +\tfrac{\MZ^2}{2}\, \delta Z^{\gamma\PZ}_{(\alpha)}\, \delta Z^{\gamma\PZ}_{(\alpha\as)} \notag \\
    &\quad + \Im\{\sigTa^{\PZ\prime}(\MZ^2)\}\, \Im\{\sigTaas^\PZ(\MZ^2)\} + \Im\{\sigTaas^{\PZ\prime}(\MZ^2)\}\, \Im\{\sigTa^\PZ(\MZ^2)\} \notag \\
    &\quad + \tfrac{2}{\MZ^2}\, \Im\{\sigTa^{\gamma\PZ}(\MZ^2)\}\, \Im\{\sigTaas^{\gamma\PZ}(\MZ^2) \}\,.
\end{align}
Here $\Sigma_{\rm T}^V$ is the transverse part of the self-energy of the vector boson $V$, and $\Re/\Im$ denote real and imaginary parts, respectively. A prime indicates a derivative with respect to the momentum-squared, $\Sigma'(p^2) \equiv \frac{\partial}{\partial (p^2)}\Sigma(p^2)$.
Furthermore, $(...)$ in the subscripts denotes the perturbative order. Note that in order to arrive at $N_f=1$ overall, the $(\alpha)$ terms are contributions \emph{without} closed fermion loops. As a result, $\Im\{\sigTa^V(M_V^2)\}$ vanishes for \mbox{$V=\PW,\PZ$}.

The weak mixing counterterm is obtained from the relation $\sw^2=1-\MW^2/\MZ^2$, leading to
\begin{align}
    \frac{\delta\sw{}_{(\alpha^2\as)}}{\sw} &=
    \frac{\cw^2}{2\sw^2}\biggl[ \frac{\delta  M^2_{\PZ(\alpha^2\as)}}{\MZ^2} - \frac{\delta  M^2_{\PW(\alpha^2\as)}}{\MW^2} \biggr] - \frac{\delta\sw{}_{(\alpha)}\, \delta\sw{}_{(\alpha\as)}}{\sw^2} \notag \\ &\quad - \frac{\delta\sw{}_{(\alpha)}\, \delta  M^2_{\PZ(\alpha\as)} + \delta\sw{}_{(\alpha\as)}\, \delta  M^2_{\PZ(\alpha)}}{\sw\MZ^2}\,.
\end{align}
For the top-quark, the OS renormalization condition reads
\begin{align}
\bigl[ Z^{t\LL}(\slashed{p}- \mt- \delta\mt)P_\LL + Z^{t\RR}(\slashed{p}- \mt - \delta\mt)P_\RR + \Sigma^\Pt(p) \bigr]_{\slashed{p}=\mu_\Pt} = 0,
\end{align}
where $P_{\LL,\RR} = (1\mp \gamma_5)/2$ and $\mu_\Pt = m_\Pt - \tfrac{i}{2} \Gamma_\Pt$ is the complex propagator pole.
This condition leads to
\begin{align}
    (Z^{t\LL}+Z^{t\RR})\delta\mt &= \mt \, \Re\bigl\{ \Sigma_\LL^\Pt(\mu^2_\Pt) + \Sigma_\RR^\Pt(\mu^2_\Pt) + 2\,\Sigma_{\rm S}^\Pt(\mu^2_\Pt) \bigr\}, \label{eq:tmass1} \\
    (Z^{t\LL}+Z^{t\RR}) \tfrac{1}{2} \Gamma_\Pt &= \mt\, \Im\bigl\{ \Sigma_\LL^\Pt(\mu^2_\Pt) + \Sigma_\RR^\Pt(\mu^2_\Pt) + 2\,\Sigma_{\rm S}^\Pt(\mu^2_\Pt) \bigr\}, \label{eq:twidth1} 
\end{align}
where we used the decomposition
\begin{align}
    \Sigma^\Pt(p) &= \slashed{p} P_\LL \, \Sigma_\LL^\Pt(p^2) + \slashed{p} P_\RR \, \Sigma_\RR^\Pt(p^2) + \mt\, \Sigma_{\rm S}^\Pt(p^2).
\end{align}
Solving eq.~\eqref{eq:twidth1} for $\Gamma_\Pt$ and inserting it into $\mu_\Pt$ in eq.~\eqref{eq:tmass1} yields, after expanding to second order in perturbation theory,
\begin{align}
    \delta\mt{}_{(\alpha\as)} &= \frac{\mt}{2}\Bigl[ \Re\bigl\{ 
    \Sigma_{\rm M(\alpha\as)}^\Pt(\mt^2) \bigr\} \notag \\
    &\qquad\quad + \mt^2\,\Im\bigl\{ 
    \Sigma_{\rm M(\alpha)}^{\Pt\prime}(\mt^2) \bigr\} \, \Im\bigl\{ 
    \Sigma_{\rm M(\as)}^\Pt(\mt^2) \bigr\} + \mt^2\,\Im\bigl\{ 
    \Sigma_{\rm M(\as)}^{\Pt\prime}(\mt^2) \bigr\} \, \Im\bigl\{ 
    \Sigma_{\rm M(\alpha)}^\Pt(\mt^2) \bigr\} \Bigr] \notag \\
    &\quad -\tfrac{1}{2}\delta\mt{}_{(\alpha)} (\delta Z^{t\LL}_{(\as)} + \delta Z^{t\RR}_{(\as)}) - \tfrac{1}{2}\delta\mt{}_{(\as)} (\delta Z^{t\LL}_{(\alpha)} + \delta Z^{t\RR}_{(\alpha)}), \label{eq:tmass2}
\end{align}
where $\Sigma_{\rm M}^\Pt(p^2) \equiv \Sigma_\LL^\Pt(p^2) + \Sigma_\RR^\Pt(p^2) + 2\,\Sigma_{\rm S}^\Pt(p^2)$. Since $\Im\{\Sigma_{\rm M(\as)}^\Pt(\mt^2)\}=0$ the terms in the second line of eq.~\eqref{eq:tmass2} vanish.  

The field renormalization counterterms for the external leptons and neutrinos are obtained from the requirement that the residue of the OS propagators is normalized to~1. Neglecting the electron, muon and neutrino masses, the resulting expressions become particularly simple:
\begin{align}
\delta Z^{f\LL} = -\Sigma^f_\LL(0), \qquad f = e,\mu,\nu_e,\nu_\mu.
\end{align}

The expression for the 3-loop charge renormalization counterterm can again be obtained from Ref.~\cite{Freitas:2002ja}, with obvious replacements, yielding
\begin{align}
    \delta Z_{e(\alpha^2\as)} &= \frac{1}{2}\Sigma^{\gamma\gamma\prime}_{\rm T(\alpha^2\as)}(0)
     - \frac{\sw}{2\cw\MZ^2} \Sigma^{\gamma\PZ}_{\rm T(\alpha^2\as)}(0) \notag \\
     &\quad + 2\,\delta Z_{e(\alpha)}\, \delta Z_{e(\alpha\as)}  +\frac 14 \delta Z^{\gamma\gamma}_{(\alpha)} \, \delta Z^{\gamma\gamma}_{(\alpha\as)}   +\frac 14 \delta Z^{\PZ\gamma}_{(\alpha)} \, \delta Z^{\PZ\gamma}_{(\alpha\as)}
   \nonumber \\
   & \begin{aligned}[b] {} \quad +\frac{\sw}{2c_W} \biggl[ &\biggl( \frac{\delta M^2_{\PZ(\alpha)}}{\MZ^2} - \frac{\delta\sw{}_{(\alpha)}}{\sw\cw^2} + \frac{1}{2}\delta Z^{\PZ\PZ}_{(\alpha)} \biggr)   \delta Z^{Z \gamma}_{(\alpha\as)} \\ &+ \biggl( \frac{\delta M^2_{\PZ(\alpha\as)}}{\MZ^2} - \frac{\delta\sw{}_{(\alpha\as)}}{\sw\cw^2} + \frac{1}{2}\delta Z^{\PZ\PZ}_{(\alpha\as)} \biggr) \delta Z^{Z \gamma}_{(\alpha)} \biggr ]. \end{aligned} \label{eq:dZe}
\end{align}

The reducible contributions at the $\mathcal{O}(\alpha^2\as)$ include 1-loop $\times$ 2-loop terms
\begin{align}
    \delta_{\rm reduc}^{(\alpha^2\as)} = 2 \, \biggl( \delta_{\rm vert}^{(\alpha)}  \delta_{\rm vert}^{(\alpha \as)} +   \delta_{\rm vert}^{(\alpha \as)}  \frac{\hat\Sigma^{\rm W}_{\rm T{(\alpha)}}(0)}{M^2_W}  +  \delta_{\rm vert}^{(\alpha)}  \frac{\hat\Sigma^{\rm W}_{\rm T{(\alpha \as)}}(0)}{M^2_W} + \frac{\hat\Sigma^{\rm W}_{\rm T{(\alpha \as)}}(0)}{M^2_W} \, \frac{\hat\Sigma^{\rm W}_{\rm T{(\alpha )}}(0)}{M^2_W} \biggr).\nonumber \\ \label{eq:vertred}
\end{align}
Here $\delta_{\rm vert}$ denotes corrections to the $\ell\nu_\ell W$ vertex relative to the Born contribution, and $\hat\Sigma_{\rm T}^{\rm W}(p^2)$ is the renormalized transverse W-boson self-energy. 

The third-order vertex correction can be decomposed as
\begin{align}
 \delta_{\rm vert}^{(\alpha^2\as)} &= 
 \delta Z_{e{(\alpha^2\as)}} 
 - \frac{\sw{}_{(\alpha^2\as)}}{\sw} + \frac{1}{2}  (\delta Z^{\rm W}_{(\alpha^2\as)} + \delta Z^{\nu\rm L}_{(\alpha^2\as)} +\delta Z^{\ell\rm L}_{(\alpha^2\as)})  \nonumber  \\
   &+ \frac 12 \frac{\sw{}_{(\alpha\as)}}{\sw} \left( 2 \frac{\delta \sw{}_{(\alpha)}}{ \sw} -2 \delta Z_{e{(\alpha)}} -\delta Z^{\rm W}_{{(\alpha)}} - \delta Z^{\nu\rm L}_{(\alpha)} -\delta Z^{\ell\rm L}_{(\alpha)} \right) \nonumber \\
&+ \frac 12 \delta Z_{e(\alpha \as)} \left( \delta Z^{\rm W}_{(\alpha)} + \delta Z^{\nu\rm L}_{(\alpha)} +\delta Z^{\ell\rm L}_{(\alpha)} -2 \frac{\delta\sw{}_{(\alpha)}}{\sw} \right) \nonumber \\
&+ \frac 14  \delta Z^{\rm W}_{(\alpha \as)}   \left( 2 \delta Z_{e(\alpha)} + \delta Z^{\nu\rm L}_{(\alpha)} +\delta Z^{\ell\rm L}_{(\alpha)} -\delta Z^{\rm W}_{(\alpha)} -2 \frac{\delta\sw{}_{(\alpha)}}{ \sw} \right)  
\nonumber \\
 &+\delta_{\rm vert}^{(\alpha^2\as),\rm bare}
 +\delta_{\rm vert}^{(\alpha)\, \rm{with}\; CT{(\alpha \as)}}+\delta_{\rm vert}^{(\alpha^2)\, \rm{with}\; CT{(\alpha_s)}}. \label{eq:vert}
\end{align}
In the last line of (\ref{eq:vert}), $\delta_{\rm vert}^{(\alpha)\, \rm{with}\; CT{(\alpha \as)}}$ is the 1-loop vertex with 2-loop counterterm insertions and $\delta_{\rm vert}^{(\alpha^2)\, \rm{with}\; CT{(\alpha_s)}}$ is the 2-loop vertex with a 1-loop counterterm insertions. Furthermore, $\delta Z^{\rm W}$ denotes the field renormalization counterterm for the W-boson, which drops out in the final result when combining the vertex corrections with self-energy corrections.
For massless electron and muon, the  $W\bar{e}\nu_e$  and  $W\bar{\mu}\nu_\mu$ vertex contributions are identical. 

The above counterterms are part of the final result for $\Delta r^{(N_f\alpha^2\as)}$, according to the general relation 
\begin{eqnarray}
    \Delta r^{(N_f\alpha^2\as)} &=& \frac{\hat\Sigma^{\rm W}_{{\rm T}{(N_f\alpha^2\as)}}(0)}{M^2_W} +  \delta_{\rm reduc}^{(N_f\alpha^2\as)} + \delta_{\rm vert}^{(N_f\alpha^2\as)} 
    + \delta_{\rm box,\rm{IR}-finite}^{(N_f\alpha^2\as)} +\delta_{\rm box,\rm{QED}}^{\rm{Fermi, subtr}}  .\label{eq:deltar1}
\end{eqnarray}
Here in addition we denote the contribution to $\Delta r$ from the IR-finite boxes involving Z and W bosons given in Fig.~\ref{fig:diags}(a)-\ref{fig:diags}(c) by $\delta_{box,\rm{IR}-finite}^{(N_f\alpha^2\as)}$, while the QED correction matched to the Fermi theory is captured by $\delta_{box,\rm{QED}}^{\rm{Fermi, subtr}}$ (see section \ref{sec:fermi}).

\bibliographystyle{elsarticle-num-ID}
\bibliography{references}

\end{document}

%% file: figures/figs_tikz.tex
\begin{figure}[tb]
\begin{tabular}{ccc}

\begin{tikzpicture}[scale=1.3]
\begin{feynman}

\vertex at (-1,0) (a);
\vertex at (0.,0) (b);
\vertex at (1.,1) (c);
\vertex at (2,1) (d);
\vertex at (2,0) (h);
\vertex at (2.5,0) (i);
\vertex at (0.25,-.25) (e);
\vertex at (.75,-.75) (f);
\vertex at (1,-1) (g);
\vertex at (2.,-1.) (j);
\vertex at (0.75,-0.35) (g1);
\vertex at (0.3,-0.7) (g2);

\diagram*{
(a) -- [thick, fermion, edge label=\( \Huge{}\)] (b) -- [thick, fermion, edge label=\( \Huge{}\)] (c) -- [thick, fermion, edge label=\( \Huge{}\)] (d), (c) -- [thick, photon, edge label=\( \Huge{}\)] (h) -- [thick, edge label=\( \Huge{}\)] (i), 
(b) -- [thick, photon, edge label=\( \Huge{}\)] (e) -- [thick, half left] (f) -- [thick, photon, edge label=\( \Huge{}\)] (g) -- [thick, fermion, edge label=\( \Huge{}\)] (j),
(g) -- [thick, edge label=\( \Huge{}\)], (i) -- [thick, fermion, edge label=\( \Huge{}\) ] (h)  -- [thick, fermion] (g), (e) -- [thick, half right] (f) , (g1) -- [thick, gluon] (g2)
};

\node at (1.3,0.3) {\small{$W$}};
\node at (.41,-.41) {\small{$g$}};
\node at (.06,-0.5) {\small{$t$}};
\node at (.7,-.62) {\small{$t$}};
\node at (-0.5,0.2) {\small{$\mu^-$}};
\node at (2.25,1) {\small{$\nu_\mu$}};
\node at (2.75,0.) {\small{$\nu_e$}};
\node at (2.25,-1) {\small{$e^-$}};
\node at (.4,0) {\small{$\gamma/Z$}}; 
 \node at (.5,-1) {\small{$\gamma/Z$}}; 
 
\end{feynman}
\end{tikzpicture}

&

\begin{tikzpicture}[scale=1.3]
\begin{feynman}

\vertex at (-1,0) (a);
\vertex at (0.,0) (b);
\vertex at (1.,1) (c);
\vertex at (2,1) (d);
\vertex at (2,0) (h);
\vertex at (2.5,0) (i);
\vertex at (0.45,-.45) (e);
\vertex at (.55,-.55) (f);
\vertex at (1,-1) (g);
\vertex at (2.,-1.) (j);
\vertex at (0.75,-0.35) (g1);
\vertex at (0.3,-0.7) (g2);

\diagram*{
(a) -- [thick, fermion, edge label=\( \Huge{}\)] (b) -- [thick, fermion, edge label=\( \Huge{}\)] (c) -- [thick, fermion, edge label=\( \Huge{}\)] (d), (c) -- [thick, photon, edge label=\( \Huge{}\)] (h) -- [thick, edge label=\( \Huge{}\)] (i), 
(b) -- [thick, photon, edge label=\( \Huge{}\)] (e), (f) -- [thick, photon, edge label=\( \Huge{}\)] (g) -- [thick, fermion, edge label=\( \Huge{}\)] (j),
(g) -- [thick, edge label=\( \Huge{}\)], (i) -- [thick, fermion, edge label=\( \Huge{}\) ] (h)  -- [thick, fermion] (g), 
};

\node at (1.3,0.3) {\small{$W$}};
\node at (-0.5,0.2) {\small{$\mu^-$}};
\node at (2.25,1) {\small{$\nu_\mu$}};
\node at (2.75,0.) {\small{$\nu_e$}};
\node at (2.25,-1) {\small{$e^-$}};
\node at (.5,-.1) {\small{$\gamma/Z$}}; 
 \node at (.45,-.9) {\small{$\gamma/Z$}}; 
  \node at (.5,-.5) {\Large{\boldmath $\otimes$}}; 
\end{feynman}
\end{tikzpicture}

&

\begin{tikzpicture}[scale=1.3]
\begin{feynman}

\vertex at (-1,0) (a);
\vertex at (0.,0) (b);
\vertex at (1.,1) (c);
\vertex at (2,1) (d);
\vertex at (2,0) (h);
\vertex at (2.5,0) (i);
\vertex at (0.55,-.55) (e);
\vertex at (.55,-.55) (f);
\vertex at (1,-1) (g);
\vertex at (2.,-1.) (j);
\vertex at (1.25,0.75) (g1);
\vertex at (1.75,0.25) (g2);
\vertex at (1.25,0.3) (h1);
\vertex at (1.75,0.7) (h2);

\diagram*{
(a) -- [thick, fermion, edge label=\( \Huge{}\)] (b) -- [thick, fermion, edge label=\( \Huge{}\)] (c) -- [thick, fermion, edge label=\( \Huge{}\)] (d), (c) -- [thick, photon, edge label=\( \Huge{}\)] (g1) -- [thick, half left] (g2) -- [thick, photon] (h) -- [thick, edge label=\( \Huge{}\)] (i), 
(b) -- [thick, photon, edge label=\( \Huge{}\)] (e), (f) -- [thick, photon, edge label=\( \Huge{}\)] (g) -- [thick, fermion, edge label=\( \Huge{}\)] (j),
(g) -- [thick, edge label=\( \Huge{}\)], (i) -- [thick, fermion, edge label=\( \Huge{}\) ] (h)  -- [thick, fermion] (g), (g1) -- [thick, half right] (g2), (h1) --[gluon] (h2)
};

\node at (1.,.75) {\small{$W$}};
\node at (1.7,0.05) {\small{$W$}};
\node at (1.1,.5) {\small{$t$}};
\node at (1.3,0.15) {\small{$t$}};
\node at (1.7,.9) {\small{$b$}};
\node at (1.9,0.4) {\small{$b$}};
\node at (1.6,.35) {\small{$g$}};
\node at (-0.5,0.2) {\small{$\mu^-$}};
\node at (2.25,1) {\small{$\nu_\mu$}};
\node at (2.75,0.) {\small{$\nu_e$}};
\node at (2.25,-1) {\small{$e^-$}};
\node at (.2,-.6) {\small{$\gamma/Z$}}; 
\end{feynman}
\end{tikzpicture}

\\[-1ex] (a) & (b) & (c) \\[1ex]

\begin{tikzpicture}[scale=.65]
\begin{feynman}

\vertex at (-0.5,0) (a);
\vertex at (1.,0) (b);
\vertex at (2.,1) (c);
\vertex at (3,2) (d);
\vertex at (4.5,2) (e);
\vertex at (5.5,2) (f);
\vertex at (2,-1) (g);
\vertex at (3,-2) (h);
\vertex at (4.5,-2) (i);
\vertex at (5.5,-2) (j);
\vertex at (-1.5,2) (k);
\vertex at (-1.5,-2) (l);

\diagram*{
(a) -- [thick, fermion] (k),
(l) -- [thick, fermion] (a) -- [thick, photon, edge label=\( \Huge{}\)] (b) -- [thick, fermion, edge label=\( \Huge{}\)] (c) -- [thick, fermion, edge label=\( \Huge{}\)] (d) -- [thick, photon, edge label=\( \Huge{}\)] (e), 
(h) -- [thick, photon, edge label=\( \Huge{}\)] (i) -- [thick, fermion, edge label=\( \Huge{}\)] (j),
(c) -- [thick, gluon, edge label=\( \Huge{}\)] (g), (d) -- [thick, fermion, edge label=\( \Huge{}\) ] (h) -- [thick, fermion, edge label=\( \Huge{}\)] (g) -- [thick, fermion, edge label=\( \Huge{}\)] (b),  (f) -- [thick, fermion, edge label=\( \Huge{}\)] (e) -- [thick, fermion, edge label=\( \Huge{}\)]
(i)  
};

\node at (0,0.3) {\small{$W$}};
\node at (2.3,0.) {\small{$g$}};
\node at (1.1,-0.6) {\small{$t$}};
\node at (3.3,0.) {\small{$t$}};
\node at (2.1,1.7) {\small{$b$}};
\node at (1.1,0.6) {\small{$b$}};
\node at (4.8,0.) {\small{$e$}};
\node at (2.1,-1.5) {\small{$t$}};
\node at (3.7,-1.6) {\small{$Z$}};
\node at (3.7,1.6) {\small{$W$}};
\node at (5,1.6) {\small{$\nu_e$}};
\node at (5,-1.6) {\small{$e$}};
\node at (-0.9,1.7) {\small{$\nu_\mu$}};
\node at (-0.9,-1.7) {\small{$\mu$}};
\end{feynman}
\end{tikzpicture}

&

\begin{tikzpicture}[scale=.65]
\begin{feynman}

\vertex at (-0.5,0) (a);
\vertex at (1.,0) (b);
\vertex at (2.,1) (c);
\vertex at (3,2) (d);
\vertex at (4.5,2) (e);
\vertex at (5.5,2) (f);
\vertex at (2,-1) (g);
\vertex at (3,-2) (h);
\vertex at (4.5,-2) (i);
\vertex at (5.5,-2) (j);
\vertex at (-1.5,2) (k);
\vertex at (-1.5,-2) (l);
\vertex at (3,0) (m);

\diagram*{
(a) -- [thick, fermion] (k),
(l) -- [thick, fermion] (a) -- [thick, photon, edge label=\( \Huge{}\)] (b) -- [thick, fermion, edge label=\( \Huge{}\)] (d) -- [thick, photon, edge label=\( \Huge{}\)] (e), 
(h) -- [thick, fermion, edge label=\( \Huge{}\)] (b), (h) -- [thick, photon, edge label=\( \Huge{}\)] (i) -- [thick, fermion, edge label=\( \Huge{}\)] (j), (d) -- [thick, fermion, edge label=\( \Huge{}\) ] (m) -- [thick, fermion, edge label=\( \Huge{}\) ] (h),  (f)  -- [thick, fermion, edge label=\( \Huge{}\)](e) -- [thick, fermion, edge label=\( \Huge{}\)]
(i)  
};

\node at (0,0.3) {\small{$W$}};

\node at (3.3,0.7) {\small{$t$}};
\node at (3.3,-0.7) {\small{$t$}};
\node at (1.8,1.3) {\small{$b$}};
\node at (4.8,0.) {\small{$e$}};
\node at (1.8,-1.3) {\small{$t$}};
\node at (3.7,-1.6) {\small{$Z$}};
\node at (3.7,1.6) {\small{$W$}};
\node at (5,1.6) {\small{$\nu_e$}};
\node at (5,-1.6) {\small{$e$}};
\node at (-0.9,1.7) {\small{$\nu_\mu$}};
\node at (-0.9,-1.7) {\small{$\mu$}};
\node at (3,0) {\Large{$\bf \times$}};
\end{feynman}
\end{tikzpicture}

&

\begin{tikzpicture}[scale=1.1]
\begin{feynman}

\vertex at (0.3,0) (a1);
\vertex at (1.,0) (a);
\vertex at (2,0) (d);
\vertex at (3,0) (f);
\vertex at (3.7,0) (a2);
\vertex at (1.5,1) (b);
\vertex at (2.5,1) (c);
\vertex at (2,-1) (e);
\vertex at (-0.2,1) (k);
\vertex at (-0.2,-1) (l);
\vertex at (4.2,1) (m);
\vertex at (4.2,-1) (n);

\diagram*{
(l) -- [thick, fermion] (a1) -- [thick, fermion] (k),
(a1) -- [thick, photon, edge label=\( \Huge{}\)] (a) -- [thick, edge label=\( \Huge{}\)] (b) -- [thick, gluon, edge label=\( \Huge{}\)] (c) -- [thick, edge label=\( \Huge{}\)] (f) -- [thick, photon, edge label=\( \Huge{}\)] (a2), (b) -- [thick, edge label=\( \Huge{}\)] (d) -- [thick, edge label=\( \Huge{}\) ] (c), 
(a) -- [thick, edge label=\( \Huge{}\)] (e) -- [thick, edge label=\( \Huge{}\)] (f), (d) -- [thick, photon, edge label=\( \Huge{ }\)] (e),
(m) -- [thick, fermion] (a2) -- [thick, fermion] (n)
};

\node at (.7,0.2) {\small{$W$}};
\node at (3.3,0.2) {\small{$W$}};
\node at (2,1.2) {\small{$g$}};
\node at (1.3,-0.5) {\small{$t$}};
\node at (2.75,-0.5) {\small{$t$}};
\node at (1.1,0.45) {\small{$b$}};
\node at (1.64,0.45) {\small{$b$}};
\node at (2.33,0.45) {\small{$b$}};
\node at (2.95,0.45) {\small{$b$}};
\node at (2.3,-0.36) {\small{$\gamma/Z$}};
\node at (0.2,0.7) {\small{$\nu_\mu$}};
\node at (0.2,-0.7) {\small{$\mu$}};
\node at (3.8,0.7) {\small{$\nu_e$}};
\node at (3.8,-0.7) {\small{$e$}};
\end{feynman}
\end{tikzpicture}

\\[-1ex] (d) & (e) & (f) \\[1ex]

\begin{tikzpicture}[scale=1.2]
\begin{feynman}

\vertex at (0.3,0) (a1);
\vertex at (1.,0) (a);
\vertex at (1.5,-1) (d);
\vertex at (3,0) (f);
\vertex at (3.7,0) (a2);
\vertex at (1.5,1) (b);
\vertex at (2.5,1) (c);
\vertex at (2.5,-1) (e);
\vertex at (1.95,0.05) (g1);
\vertex at (2.02,-0.07) (g2);
\vertex at (-0.2,1) (k);
\vertex at (-0.2,-1) (l);
\vertex at (4.2,1) (m);
\vertex at (4.2,-1) (n);

\diagram*{
(l) -- [thick, fermion] (a1) -- [thick, fermion] (k),
(a1) -- [thick, photon, edge label=\( \Huge{}\)] (a) -- [thick, edge label=\( \Huge{}\)] (b) -- [thick, gluon, edge label=\( \Huge{}\)] (c) -- [thick, edge label=\( \Huge{}\)] (f) -- [thick, photon, edge label=\( \Huge{}\)] (a2), (b) -- [thick, edge label=\( \Huge{}\)] (g1), (g2) -- [thick] (e)  -- [thick, edge label=\( \Huge{}\) ] (f),  (a) -- [thick, edge label=\( \Huge{}\)]
(d) -- [thick, edge label=\( \Huge{}\)] (c) -- [thick, edge label=\( \Huge{}\)] (f), (d) -- [thick, photon, edge label=\( \Huge{}\)] (e),
(m) -- [thick, fermion] (a2) -- [thick, fermion] (n) 
};

\node at (.7,0.2) {\small{$W$}};
\node at (3.3,0.2) {\small{$W$}};
\node at (2,1.2) {\small{$g$}};
\node at (1.27,-0.3) {\small{$t$}};
\node at (2.73,-0.27) {\small{$b$}};
\node at (2.7,0.33) {\small{$t$}};
\node at (1.27,0.3) {\small{$b$}};
\node at (1.83,0.6) {\small{$b$}};
\node at (1.75,-.7) {\small{$t$}};
\node at (2,-1.2) {\small{$\gamma/Z$}};
\node at (0.2,0.7) {\small{$\nu_\mu$}};
\node at (0.2,-0.7) {\small{$\mu$}};
\node at (3.8,0.7) {\small{$\nu_e$}};
\node at (3.8,-0.7) {\small{$e$}};
\end{feynman}
\end{tikzpicture}

&

\begin{tikzpicture}[scale=1.5]
\begin{feynman}

\vertex at (0,0) (a);
\vertex at (0.5,0) (b);
\vertex at (1.,1) (c);
\vertex at (2,1) (d);
\vertex at (1.5,0) (h);
\vertex at (2.5,0) (i);
\vertex at (0.5,-.5) (e);
\vertex at (1.2,-.5) (f);
\vertex at (2.7,-.8) (g);
\vertex at (2.7,.8) (j);
\vertex at (0.8,-0.78) (g1);
\vertex at (0.8,-0.2) (g2);
\vertex at (2.2,-0.5) (g3);

\diagram*{
(a) -- [thick, fermion, edge label=\( \Huge{}\)] (h),  (h) -- [thick, edge label=\( \Huge{}\)] (i), 
(b) -- [thick, photon, quarter right] (g3), (h) -- [thick, fermion, edge label=\( \Huge{}\)] (j),
(h) -- [thick, fermion, edge label=\( \Huge{}\)] (g), (i) -- [thick, fermion, edge label=\( \Huge{}\) ] (h)  
};

\node at (0.,0.2) {\small{$\mu^-$}};
\node at (2.6,0.5) {\small{$\nu_\mu$}};
\node at (2.75,0.) {\small{$\nu_e$}};
\node at (2.6,-.5) {\small{$e^-$}};
\node at (1.1,-.8) {\small{$\gamma$}}; 
\node at (1.5,0){\rule{1.5ex}{1.5ex}}; 
\end{feynman}
\end{tikzpicture}

&

\begin{tikzpicture}[scale=1.3]
\begin{feynman}

\vertex at (-1,0) (a);
\vertex at (0.,0) (b);
\vertex at (1.,1) (c);
\vertex at (2,1) (d);
\vertex at (1.5,0) (h);
\vertex at (2.5,0) (i);
\vertex at (0.5,-.5) (e);
\vertex at (1.2,-.5) (f);
\vertex at (2.7,-.8) (g);
\vertex at (2.7,.8) (j);
\vertex at (0.8,-0.78) (g1);
\vertex at (0.8,-0.2) (g2);
\vertex at (1.8,-0.23) (g3);

\diagram*{
(a) -- [thick, fermion, edge label=\( \Huge{}\)] (h),  (h) -- [thick, edge label=\( \Huge{}\)] (i), 
(b) -- [thick, photon, edge label=\( \Huge{}\)] (e) -- [thick, half left] (f) -- [thick, photon, edge label=\( \Huge{}\)] (g3), (h) -- [thick, fermion, edge label=\( \Huge{}\)] (j),
(g) -- [thick, edge label=\( \Huge{}\)], (i) -- [thick, fermion, edge label=\( \Huge{}\) ] (h)  -- [thick, fermion] (g), (e) -- [thick, half right] (f) , (g1) -- [thick, gluon] (g2)
};

\node at (0.96,-.49) {\small{$g$}};
\node at (1.1,-0.85) {\small{$t$}};
\node at (.55,-.85) {\small{$t$}};
\node at (-0.5,0.2) {\small{$\mu^-$}};
\node at (2.25,0.8) {\small{$\nu_\mu$}};
\node at (2.75,0.) {\small{$\nu_e$}};
\node at (2.25,-.8) {\small{$e^-$}};
\node at (0.1,-.4) {\small{$\gamma$}}; 
\node at (1.6,-.5) {\small{$\gamma$}}; 
\node at (1.5,0){\rule{1.5ex}{1.5ex}};
\end{feynman}
\end{tikzpicture}

\\[-1ex] (g) & (h) & (i)

\end{tabular}
\caption{Sample diagrams for ${\cal O}(\alpha^2\as)$ corrections to muon decay in the full SM (a--g), and for the matching to the Fermi model (h,i). The symbols $\times$ and $\otimes$ denote ${\cal O}(\as)$ and ${\cal O}(\alpha\as)$ counterterms, respectively, whereas the black squares represent effective four-fermion interactions in the Fermi model.}
\label{fig:diags}
\end{figure}
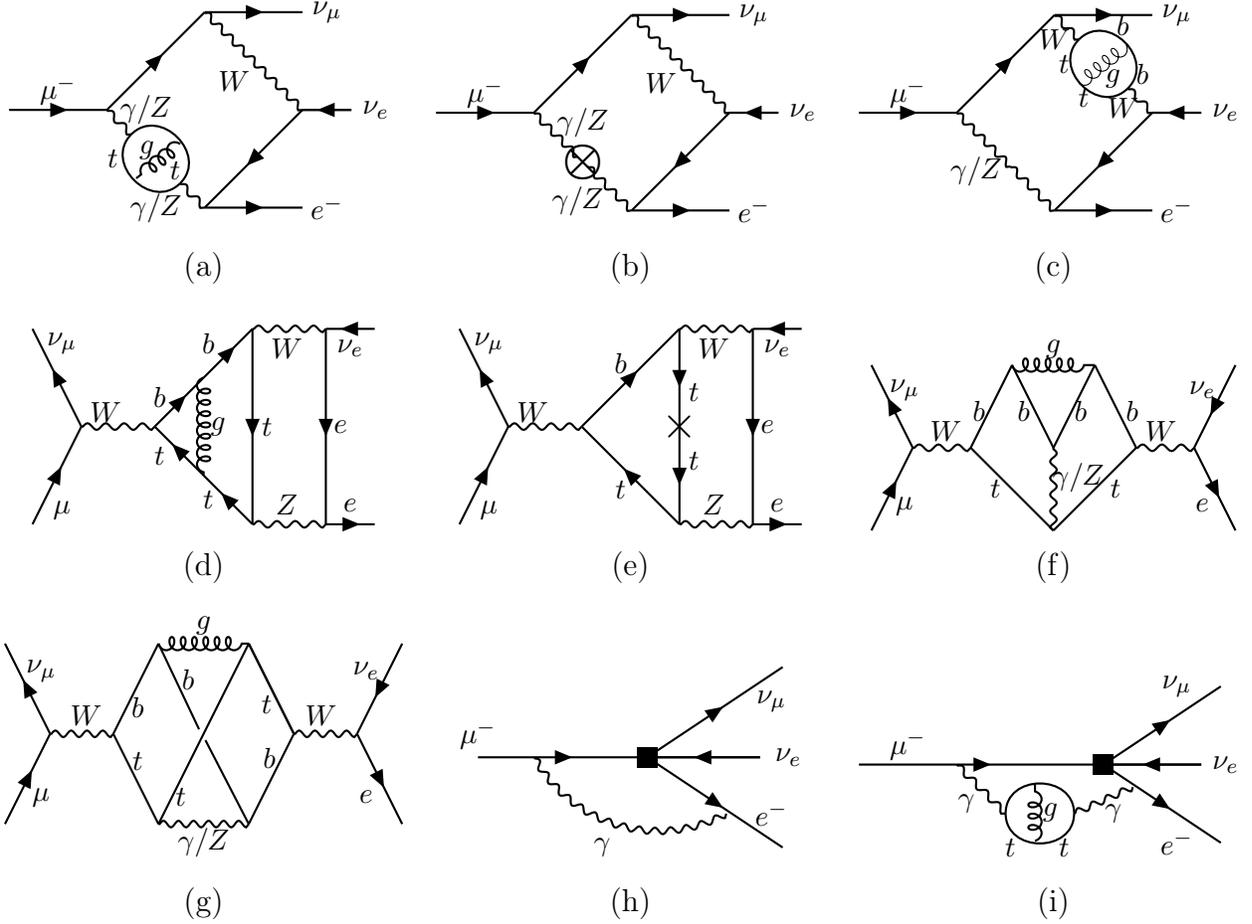